\documentclass[twocolumn, pra]{revtex4-1}

\usepackage{booktabs}
\usepackage{palatino}
\usepackage{verbatim}
\usepackage{amsmath,amsthm}
\usepackage{amssymb}
\usepackage{graphicx}
\usepackage{color}
\usepackage{tikz}
\usetikzlibrary{backgrounds,fit,decorations.pathreplacing,calc,,shadows}

\usepackage{natbib}
\usepackage{epsfig}

\def\ket#1{|{#1}\rangle}

\newcommand{\ketbra}[2]{|#1\rangle\langle#2|}

\newcommand{\trace}{\operatorname{Tr}}

\begin{document}

\title{Quantum Channel Negativity as a Measure of System-Bath Coupling and Correlation}

\author{James M. McCracken}
\email{james.mccracken@nrl.navy.mil}
\affiliation{U.S. Naval Research Laboratory, Code 5540, Washington, DC 20375}
\date{\today}

\begin{abstract}
Complete positivity is a ubiquitous assumption in the study of quantum systems interacting with the environment, but the lack of complete positivity of a quantum evolution (called the ``negativity'') can be used as a measure of the system-bath coupling and correlation.  The negativity can be computed from the Choi representation of a channel, is always defined and bounded, and contains some information about environmentally induced noise in a quantum system.   
\end{abstract}
\maketitle
 
\section{Introduction}
Complete positivity has become an ingrained part of the modern study of open quantum systems, but dynamics that are not completely positive have recently garnered much interest in the study of general quantum evolutions\cite{Pechukas1,Alicki95,Pechukas2,Sud05,Rodríguez07} (and references therein).  The question of dynamics that are ``accessible'' (and not completely positive) has recently been addressed \cite{Terno08}, and this work will ask the question of what (if any) information about the bath can be gleaned from such dynamics.  

In this work, a non-completely positive quantum evolution is called ``negative''.  A composite quantum system is a quantum system under the control of the experimenter (called the ``reduced system'') along with the other quantum systems inaccessible to the experimenter that may still influence the dynamics of the reduced system (called the ``bath'', ``environment'', ``reservoir'', etc).  It will be shown that a measurement of the negativity (defined below) will give an experimenter some understanding of the coupling and correlations between the reduced system and bath.  The ``correlation'' is defined as the initial correlation between the states of the reduced system and bath, and the term will usually be used to refer to the assignment (or sharp) operator that is defined below.  The ``coupling'' is defined as the interaction between the reduced system and the bath, and the term will be used to refer to the composite dynamics.

The negativity is a straightforward calculation using the Choi representation of a channel, which is part of all quantum process tomography experiments.  These kinds of experiments are gaining popularity in the effort to engineer new quantum technologies, including quantum information processors.  An experimenter can quickly calculate the channel negativity as he finds the superoperator describing the gate he is implementing.

Such experiments have already measured non-zero negativities.  Consider the experiment conducted by Weinstein et al.\ \cite{Cory2004}, in which process tomography is performed on a three-qubit NMR quantum information processor.  The process tomography of this experiment leads to a ``non-completely positive superoperator'', i.e.\ a negative channel, which is characterized using a ``positivity''.  It can be shown that the positivity $\varrho$ is related to the channel negativity (introduced below) as
\begin{equation}
\eta  = \frac{1 - \varrho}{2 - \varrho}\;\;.
\end{equation}
The positivity of the experimental data in this NMR experiment (which was $\varrho=0.60$) corresponds to a negativity of $\eta\approx 0.29$.  These authors have experimental evidence of their system-bath coupling and correlation, and attempting to match this negativity measurement to theoretical models (and numerical simulations) will lead to a better understanding of the open system dynamics of their experiment.  

The negativity in a tomography experiment can grant the experimenter knowledge about channel parameters that cannot be measured directly. It condenses that information into a single parameter that can be useful for understanding how changes to the channel parameters change the channel. For example, the negativity can be plotted to find relationships with the channel parameters. 

\section{Complete Positivity Assumptions}
It is difficult to discuss the measurement of channel negativity in tomography experiments without discussing complete positivity.  A measurement of a non-zero negativity implies reduced quantum dynamics that at least appear to be negative.  The measurement of negativity in tomography experiments is an important part of understanding how and why reduced dynamics can appear negative, and if such dynamics are not completely positive under ideal conditions (e.g.\ if the tomography was conducted perfectly without noisy measurements and preparations).  As such, before the discussion turns to the negativity, it is a good idea to discuss why complete positivity is the standard assumption in quantum information.  

If a map is a valid quantum map, then it must take valid density matrices to valid density matrices.  A trivial extension of the map is physically reasonable and must result in a valid quantum map, i.e.\ the trivial extension of the quantum map must also take valid density matrices to valid density matrices.  Therefore, the quantum map must be completely positive.  This argument is called the ``total domain'' argument for complete positivity.  The total domain argument has recently been applied to entanglement between quantum subsystems \cite{Benatti2005}.  This new formulation of the total domain argument is often used to justify complete positivity as a physical requirement based on entanglement \cite{Alicki2007}.  

Notice, however, that it is possible to define a positivity domain as the domain of states in which a map $\Gamma$ will be positive.  On the positivity domain, $\Gamma$ will take valid initial states to valid final states.  Such a requirement is identical in spirit to the requirement of complete positivity (given the total domain argument) except that it is not extended to states which are not actually created in the lab. 

Other authors argue that the reduced system and bath must initially be uncorrelated.  Such a situation will always lead to completely positive reduced dynamics.  This argument is referred to as the ``product state'' argument for complete positivity.  

Many authors have argued that reduced dynamics do not need to be completely positive.  There are flaws in both the total domain and the product state arguments for the physical motivation of complete positivity.  However, this issue appears to be unsettled in the literature.  Arguments against complete positivity appeared in 1994 in a paper by Pechukas \cite{Pechukas1}, which led to a subsequent rebuttal/comment chain with Alicki \cite{Alicki95}.  In 2005 Sudarshan again pointed out the problem with only focusing on completely positive maps \cite{Sud05}.  Despite these papers, the idea of complete positivity as a physical requirement is still a very big part of the quantum information community.  It is often considered as part of the definition of a quantum channel \cite{nielsen00}.

The question of whether or not reduced dynamics must have completely positive descriptions is still an open question.  It is possible, however, to write down reduced dynamics (which can be found using standard tomography techniques) which appear negative.  Examples of negative channels can be found in \cite{rodriguez08}, and a few will be discussed below.

\section{Choi Representation}
A quantum channel $\varepsilon$ is defined in the open systems setting as 
\begin{equation}
\label{eq:channel}
\varepsilon(\rho) = \trace_B\left(U \rho^\sharp U^\dagger\right)\;\;,
\end{equation} 
where $\rho$ is the initial state of the reduced system, $U$ is the unitary evolution of the composite system, and $\sharp$ is called the ``assignment map'' (or ``sharp operator'' or ``extension map'').  The state $\rho$ resides in the set of valid states on the Hilbert space accessible to the experimenter in the lab, $\mathcal{H}^S$, and the evolution of the reduced system is found by ``tracing out'' the bath from the joint evolution of the reduced system and the bath \cite{breuer07}; i.e.\ $U$ resides in the set of bounded operators on the Hilbert space $\mathcal{H}^{SB}=\mathcal{H}^S\otimes\mathcal{H}^B$ where $\mathcal{H}^B$ is the Hilbert space of the bath.  The partial trace operation, $\trace_B$, is an operator that allows expectation values of observables in the reduced system to be consistent with trivial extensions into a higher dimensional Hilbert space \cite{cohen77}.  The sharp operation (or ``assignment map'') is an operation that injects the initial state of the reduced system into the higher dimensional Hilbert space of the composite system \cite{Pechukas1,Pechukas2,rodriguez10} and might only be defined on a subset of states in $\mathcal{H}^S$.  The channel should take valid quantum states to valid quantum states, hence $\varepsilon$ is to be positive (on some domain of states called the ``positivity domain''), Hermiticity-preserving, and consistent, i.e.\ $\trace_B\left(\rho^\sharp\right) = \rho$.  It is assumed here that $\mathcal{H}^S$, $\mathcal{H}^B$, and $\mathcal{H}^{SB}$ are all finite dimensional.  More information about the mathematical structure of quantum information channels in the open system settings can be found in \cite{rodriguez10,breuer07}.

Every channel $\varepsilon$ will have a Choi representation $\mathbf{C}$ given as
\begin{equation}
\mathbf{C} = \sum_{ij} E_{ij}\otimes \varepsilon\left(E_{ij}\right) \;\;,
\end{equation}  
where $E_{ij}$ is a matrix with the same dimensions as the reduced system that has a 1 at the $ij$th position and 0 everywhere else.  The matrix $\mathbf{C}$ is commonly called ``Choi's matrix'' \cite{Choi75}.  For a single qubit channel, Choi's matrix takes a simple block form, i.e.
\begin{equation}
\label{eq:Choi}
\mathbf{C} = \left(\begin{array}{c|c}
\varepsilon\left(\ketbra{0}{0}\right)&\varepsilon\left(\ketbra{0}{1}\right)\\
\hline
\varepsilon\left(\ketbra{1}{0}\right)&\varepsilon\left(\ketbra{1}{1}\right)\\
\end{array}\right)\;\;.
\end{equation}
The assumed linearity of the channel allows the off-diagonal blocks to be found using single qubit process tomography \cite{nielsen00}; e.g.\
\begin{equation}
\ketbra{0}{1} = \vec{r}\cdot\vec{\tau}\;\;,
\end{equation}
with the complex vector $\vec{r}=\{(1-i)/2,1,i,(1-i)/2\}$ and the vector of states (or ``tomography vector'') $\vec{\tau}=\{\ketbra{0}{0},\ketbra{+}{+},\ketbra{+_i}{+_i},\ketbra{1}{1}\}$ given $\ket{\pm}=(\ket{0}\pm\ket{1})/\sqrt{2}$ and $\ket{\pm_i}=(\ket{0}\pm i\ket{1})/\sqrt{2}$.  The linearity of $\varepsilon$ implies
\begin{equation}
\varepsilon\left(\ketbra{0}{1}\right) = \vec{r}\cdot\varepsilon\left(\vec{\tau}\right)\;\;;
\end{equation}
i.e.\ the matrix $\mathbf{C}$ is defined by the tomographic characterization of a channel.

\section{Definition}
The channel negativity is defined as 
\begin{equation}
\eta \equiv \frac{\sum_i |\lambda_i|}{\sum_j |\lambda_j|} = \frac{1}{2}\left(1-\frac{\trace\left(\mathbf{C}\right)}{||\mathbf{C}||_1}\right)\;\;,
\end{equation} 
where $\lambda$ is an eigenvalue of $\mathbf{C}$, $\lambda_i<0\;\forall i$, and $||\mathbf{C}||_1\equiv\trace\left(\sqrt{\mathbf{C}^\dagger\mathbf{C}}\right)$ is the trace norm of $\mathbf{C}$.  Notice, $\sum_j |\lambda_j|=\trace\left(\mathbf{C}\right)$ if and only if the negativity is zero.

Notice, the trace condition of $\mathbf{C}$ implies
\begin{equation}
\sum_j |\Lambda_j| \ge \trace\left(\mathbf{C}\right) > 0\;\;.
\end{equation}
Hence, the channel negativity $\eta$ is never undefined and $\eta\in[0,1/2)$ with the vanishing negativity occurring if and only if the channel is completely positive.

The additivity of the channel negativity is still an open question.  As such, this article will not discuss composite channels. 

\section{Examples}
Before the discussion turns to the possible utility of the channel negativity, consider a few simple example calculations.  The negativity requires a process tomography experiment to find the Choi representation of the channel, and in turn, the channel definition requires a sharp operation defined on the tomography vector of states used in the process tomography experiment.  The sharp operations defined in the examples are given without motivation but discussion about physical realizations of different sharp operators (including the ones used in this article) can be found in \cite{rodriguez10,rodriguez08}.

\subsection{Root Swap Gate}
Consider an example of two qubits.  One qubit will be the reduced system and the other will act as the bath.  If the composite dynamics are defined as the root swap gate
\begin{equation}
U_{\sqrt{Sw}} = \frac{1}{\sqrt{2}}\begin{pmatrix}
\sqrt{2}&0&0&0\\
0&1&i&0\\
0&i&1&0\\
0&0&0&\sqrt{2}
\end{pmatrix}\;\;,
\end{equation}
then the channel becomes
\begin{equation}
\varepsilon(\rho) = \trace_B\left(U_{\sqrt{Sw}} \rho^\sharp U_{\sqrt{Sw}}^\dagger\right)\;\;.
\end{equation} 
Define the sharp operation on the canonical tomography vector $\vec{\tau}$ (introduced above) as
\begin{equation}
\label{eqn:sharpex}
\tau_i^\sharp = \tau_i\otimes \left(H\tau_i H^\dagger\right)
\end{equation}
where $H$ is the Hadamard gate \cite{nielsen00}.  This sharp operation acts on the single qubit reduced system initial states and yields two qubit composite system initial states as follows:
\begin{eqnarray*}
\left(\ketbra{0}{0}\right)^\sharp &=& \ketbra{0+}{0+}\\
\left(\ketbra{+}{+}\right)^\sharp &=& \ketbra{+0}{+0}\\
\left(\ketbra{+_i}{+_i}\right)^\sharp &=& \ketbra{+_{i}-_i}{+_{i}-_i}\\
\left(\ketbra{1}{1}\right)^\sharp &=& \ketbra{1-}{1-}\;\;.
\end{eqnarray*}

Process tomography of this channel yields a Choi representation of
\begin{equation}
\mathbf{C}_{\sqrt{Sw}} = \begin{pmatrix}
 \frac{3}{4} & -\frac{i}{2 \sqrt{2}} & \frac{1}{4} & \frac{\frac{1}{2}+\frac{i}{2}}{\sqrt{2}} \\
 \frac{i}{2 \sqrt{2}} & \frac{1}{4} & \frac{\frac{1}{2}-\frac{i}{2}}{\sqrt{2}} & -\frac{1}{4} \\
 \frac{1}{4} & \frac{\frac{1}{2}+\frac{i}{2}}{\sqrt{2}} & \frac{1}{4} & -\frac{i}{2 \sqrt{2}} \\
 \frac{\frac{1}{2}-\frac{i}{2}}{\sqrt{2}} & -\frac{1}{4} & \frac{i}{2 \sqrt{2}} & \frac{3}{4}
\end{pmatrix}\;\;,
\end{equation}
and a channel negativity of $\eta_{\sqrt{Sw}} \approx 0.149$.

\subsection{CZ Gate}
If the composite dynamics are defined as a controlled phase gate
\begin{equation}
CZ = \begin{pmatrix}
1&0&0&0\\
0&1&0&0\\
0&0&1&0\\
0&0&0&-1
\end{pmatrix}
\end{equation}
then the channel becomes
\begin{equation}
\varepsilon(\rho) = \trace_B\left(CZ \rho^\sharp CZ^\dagger\right)\;\;.
\end{equation} 
A Choi representation of this channel can be found using the sharp operator from the previous example; i.e.\
\begin{equation}
\label{eqn:Ccz}
\mathbf{C}_{CZ} = \begin{pmatrix}
 \frac{1}{2} & \frac{1}{2} & \frac{1}{2} & -\frac{1}{2}-\frac{i}{2} \\
 \frac{1}{2} & \frac{1}{2} & -\frac{1}{2}-\frac{i}{2} & -\frac{1}{2} \\
 \frac{1}{2} & -\frac{1}{2}+\frac{i}{2} & \frac{1}{2} & \frac{1}{2} \\
 -\frac{1}{2}+\frac{i}{2} & -\frac{1}{2} & \frac{1}{2} & \frac{1}{2}
\end{pmatrix}\;\;.
\end{equation}
The negativity of this channel is $\eta_{CZ}\approx 0.167$.

In both of these examples, the channel negativity is fixed because the composite dynamics (i.e.\ the ``coupling'') and sharp operations (i.e.\ the ``correlation'') are fixed.  

\subsection{Rabi Channel}
\label{sec:Rabi}
A slightly more complicated example of the negativity can be given as follows: Suppose two Rabi atoms are coupled in some way that will be represented by $H_C$ defined below.  One atom will act as our reduced system and the other will be the bath.  The physical mechanism for their coupling will not be important to this discussion, but the coupling terms in the Hamiltonian will change the behaviour of the negativity.  The channel negativity of the single qubit channel in this example will be a single real number that yields information about the system-bath coupling (the correlation will be fixed in this example).  

The channel will be governed by the Hamiltonian
\begin{equation}
H_u = \frac{\hbar}{2}\begin{pmatrix}
-\nu_s & \Omega_s\\
\Omega_s & \nu_s
\end{pmatrix}\otimes I + I\otimes \frac{\hbar}{2}\begin{pmatrix}
-\nu_b & \Omega_b\\
\Omega_b & \nu_b
\end{pmatrix} + H_C\;\;,
\end{equation}
where $\nu$ is the detuning, $\Omega$ is the (real) Rabi frequency, the subscript $s$ ($b$) indicates a term describing the reduced system (bath), $I$ is the identity operator and $H_C$ describes the coupling between the two atoms.  The coupling Hamiltonian acts on both atoms and can be written in general as
\begin{equation}
H_C = \sum_{ij} k_{ij} \sigma_i\otimes \sigma_j
\end{equation}
where $k_{ij}$ are constants and the $\sigma_i$ operators are the standard Pauli operators.  For simplicity, consider 
\begin{equation}
H_C = k_z \sigma_3\otimes\sigma_3
\end{equation}
where $k_z=k_{33}$ is some positive real number.  Assume that the atoms are identical and subject to the same classical field.  These assumptions lead to
\begin{equation}
\label{eqn:hu}
H_u^\prime = H_q\otimes I + I\otimes H_q + k_z\sigma_3\otimes\sigma_3\;\;,
\end{equation}
where $H_q$ is given as
\begin{equation}
\label{eqn:hq}
H_q = \frac{\hbar}{2}\begin{pmatrix}
-\nu & \Omega\\
\Omega & \nu
\end{pmatrix}\;\;.
\end{equation}
This Hamiltonian leads to the composite dynamics of
\begin{equation}
U = \exp{\left\{-\frac{i}{\hbar}t H_u^\prime\right\}}\;\;.
\end{equation}

The composite dynamics are calculated in the ``bare basis'' of the Hamiltonian (i.e.\ the basis in which the Hamiltonian is written in Eqn.\ \ref{eqn:hu}) in all of the calculations that follow.  This is an important point to remember.  Rabi models are typically used to describe evolution in a ``dressed basis'', which is the diagonal basis of the Rabi Hamiltonian.  In this work, however, everything will be done in the bare basis to help keep a clear interpretation of the results.  This basis choice for the composite dynamics implies that the canonical tomography vector, the sharp operation, and the partial trace are all likewise defined using the bare basis.

The next step to defining this channel is to define the sharp operation and the tomography vector.  The sharp operation will be defined as
\begin{equation}
\tau_i^\sharp = \tau_i\otimes\left(R(\phi)\tau_iR(\phi)^\dagger\right)\;\;,
\end{equation}
where 
\begin{equation}
R(\phi) = \begin{pmatrix}
\cos{\phi}&-\sin{\phi}\\
\sin{\phi}&\cos{\phi}
\end{pmatrix}
\end{equation}
with $\phi\in\mathbb{R}$ and $\vec{\tau}$ is given above.  This sharp operation is sufficiently general to study its effects the negativity, and it stills meets nessecary the requirements.  

The reduced dynamics for a reduced state $\tau_i$ are
\begin{equation}
\varepsilon\left(\tau_i\right) = \trace_B\left(U \tau_i^\sharp U^\dagger\right)\;\;,
\end{equation}
and the Choi representation of the single qubit channel is 
\begin{equation}
\mathbf{C} = \sum_{ij} \ketbra{i}{j}\otimes \varepsilon\left(\ketbra{i}{j}\right)
\end{equation}
where $\varepsilon\left(\ketbra{i}{j}\right)$ found from a process tomography experiment using the tomography vector $\vec{\tau}$ from the definition of the sharp operation.

This channel is not as simple as the previous examples.  The channel is dependent on five separate parameters: the detuning of the two atoms $\nu$, the Rabi frequency of the two atoms $\Omega$, the coupling between the two atoms $k_z$, the initial rotation of the bath atom $\phi$, and time $t$.  Such a 5-dimensional parameter space can make the Choi representation of the channel difficult to interpret directly.  However, the behaviour of the channel negativity can be investigated to draw a some conclusions about the channel.  

This example can be simplified with a few assumptions about the system.  The field is assumed to be resonant with the atoms, i.e.\ $\nu=0$. All of the parameters are scaled such that $\Omega=1$ with everything in units of $\hbar=1$.  The initial rotation of the bath atom will be fixed by a Hadamard rotation rather than $R(\phi)$, i.e.\ the sharp operation will be given by Eqn.\ \ref{eqn:sharpex}.  These assumptions will reduce the 5-dimensional parameter space to a much simpler 2-dimensional parameter space of $(k_z,t)$.

\begin{figure}[th]
\includegraphics[scale=0.52]{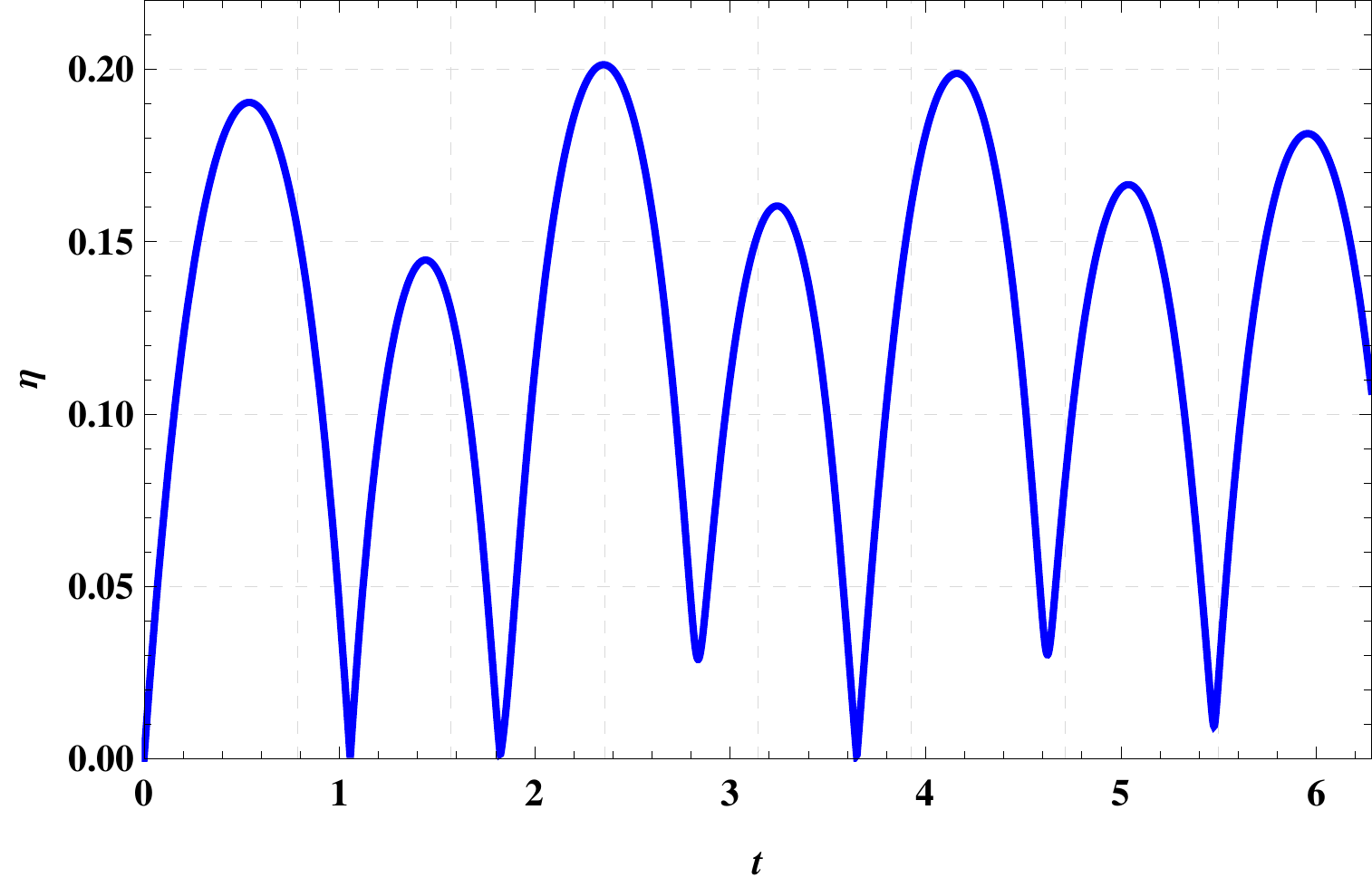}
\caption{The negativity $\eta$ only depends on the elapsed time $t$ when the coupling constant $k_z$ is fixed.  This plot is for a fixed coupling coefficient of $k_z=\pi/2$ (and given the assumptions discussed in the text).  The negativity is unitless by construction.  The time $t$ is plotted as unitless (see the text for a discussion of the units for this plot).} 
\label{fig:plot1}
\end{figure}

\begin{figure}[th]
\includegraphics[scale=0.52]{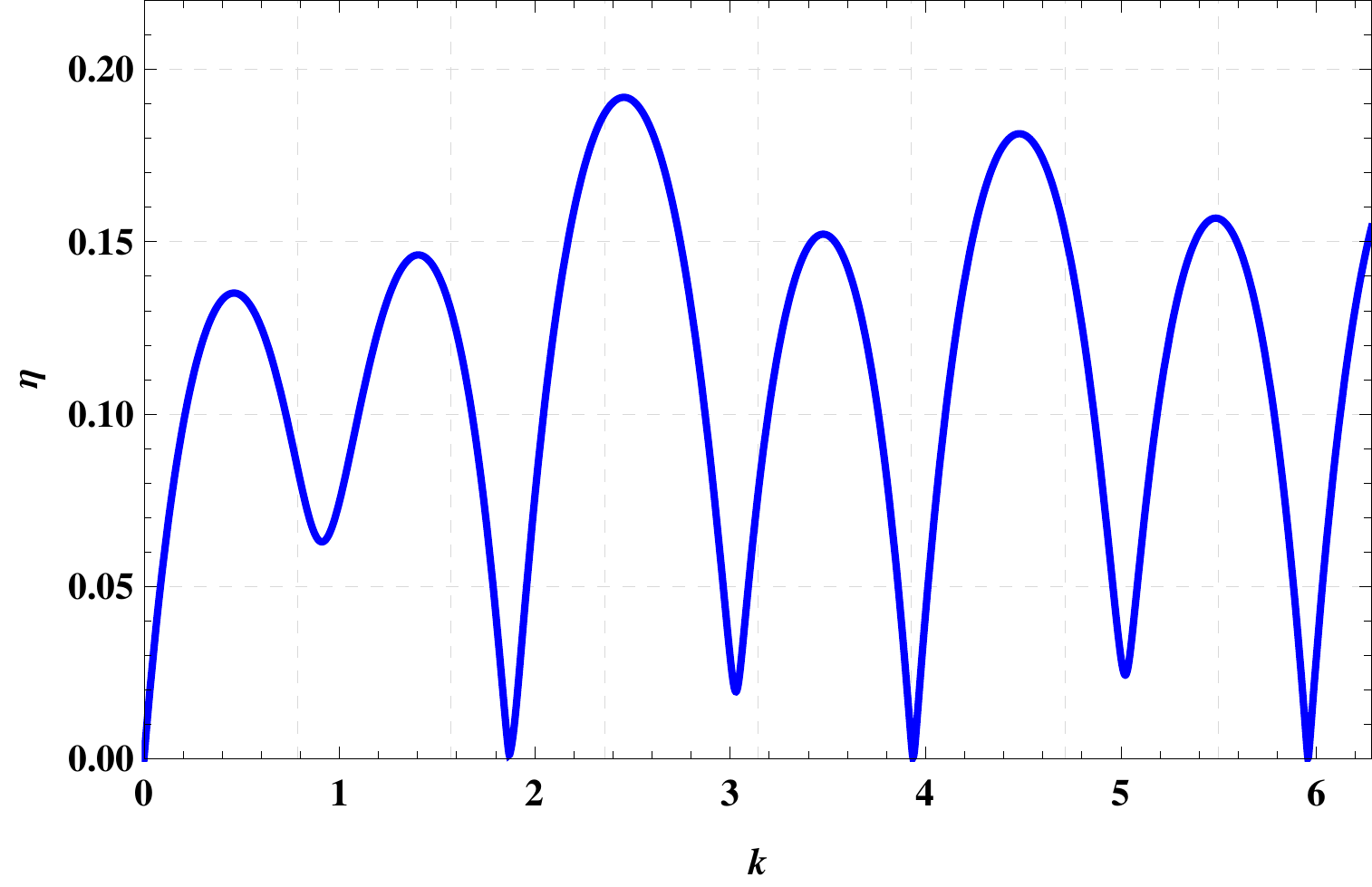}
\caption{The negativity $\eta$ only depends on  the coupling constant $k_z$ when the elapsed time $t$ is fixed.  This plot is for a fixed time of $t=\pi/2$.  (See the text for a discussion of the other assumptions.)  The coupling $k_z$ is plotted as unitless (see the text for a discussion of the units for this plot).}
\label{fig:IIplot2}
\end{figure}

Fig.\ \ref{fig:plot1} is the behaviour of the negativity as a function of time if the coupling coefficient is fixed at $k_z=\pi/2$.  
Fig.\ \ref{fig:IIplot2} is the behavior of the negativity as a function of the coupling coefficient if the time is fixed at $t=\pi/2$.  
Both plots show the channel negativity $\eta$ calculated with one parameter (either $t$ or $k_z$) over the range $[0,2\pi]$ while holding the other parameter fixed.  Notice that the negativity is only zero (i.e.\ the channel is only completely positive) at a small number of fixed points.

\begin{figure}[th]
\includegraphics[scale=0.55]{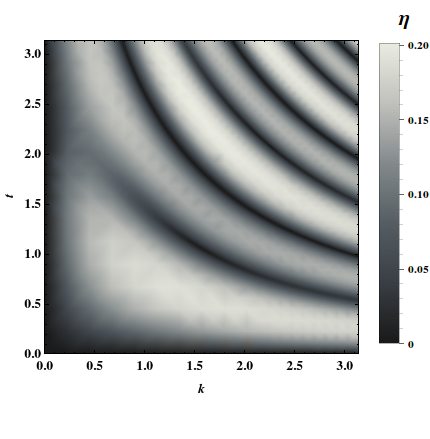}
\caption{The negativity $\eta$ of the example Rabi channel in the text depends on both the coupling constant $k_z$ and the elapsed time $t$.  Notice that the channel is almost always negative, and the negativity appears cyclic.  The text discusses the assumptions and units used to produce this plot.}
\label{fig:IIIplot3}
\end{figure}

The negativity can be plotted as a function of both the coupling coefficient and time for a visualization of the 2-dimensional parameter space over the range $[0,\pi]$.  Fig.\ \ref{fig:IIIplot3} is that plot.

This more complicated example shows that the negativity is dependent on the physical parameters of the Hamiltonian.  As such, a measurement of the negativity (through a tomography experiment) will yield information about parameters that might be inaccessible directly, e.g.\ the coupling $k_z$ between the reduced system and bath qubits in this Rabi channel.

\section{Probing the Bath}

\subsection{Coupling Alone}
Consider again two qubits.  The composite dynamics will be described by the general two qubit rotation
\begin{equation}
U_\theta = \begin{pmatrix}
1&0&0&0\\
0&\cos\theta&\sin\theta&0\\
0&-\sin\theta&\cos\theta&0\\
0&0&0&1
\end{pmatrix}\;\;.
\end{equation}
The sharp operation will be defined on the canonical tomography vector and will take the same form as before (i.e.\ $\tau_{i}^\sharp = \tau_i\otimes \left(H\tau_{i} H^\dagger\right)$), which leads to a Choi representation $\mathbf{C}_\theta$ of the channel $\varepsilon(\tau_i)=\trace_B\left(U_\theta\tau_{i}^\sharp U_\theta^\dagger\right)$.  (The Choi representation of this channel is complicated and does not need to be written down directly for this discussion.  However, interested readers can find it in \cite{thesis} along with the details of all the example calculations presented here.) 

\begin{figure}[th]
\includegraphics[scale=0.52]{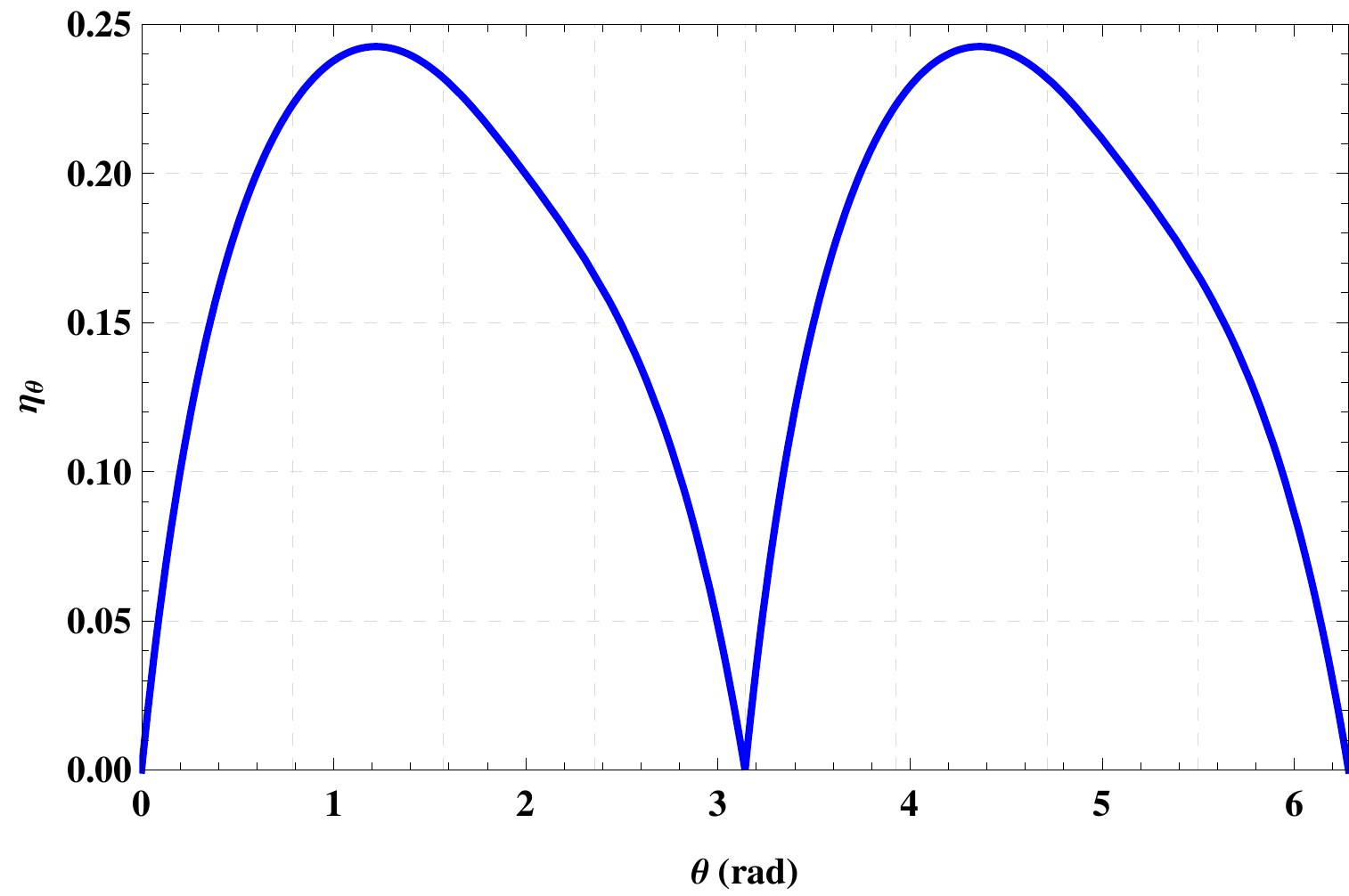}
\caption{The negativity $\eta_\theta$ can be plotted as a function of $\theta$ to show the dependency of the negativity on $U_\theta$, despite the analytical difficulties of finding the spectrum of $C_\theta$.  See the text for definitions of $U_\theta$ and $C_\theta$.  The negativity is unitless and $\theta$ has units of radians.}
\label{fig:etatheta}
\end{figure}

Notice, $\theta=0$ and $\theta=2\pi$ lead to $U_\theta=I$ where $I$ is the two qubit identity operator.  Define, $\eta_\theta$ to be the negativity of the channel represented by $\mathbf{C}_\theta$.  If $\theta\in[0,2\pi]$, then there are three points where $\eta_\theta = 0$: $\theta=0$, $\theta=2\pi$, and $\theta=\pi$.  The negativity can be plotted as function of $\theta$ (see Fig.\ \ref{fig:etatheta}) to reveal a maximum negativity of $\eta_\theta\approx 0.24$ around the points $\theta=\pi/3$ and $\theta=4\pi/3$.
The negativity $\eta_\theta$ is a function of $\theta$ and can, in principle, be used to gain information about $\theta$.  

The negativity can be measured and if the above theoretical definition of the channel is assumed to be true, then $\theta$ can simply be read off Fig.\ \ref{fig:etatheta}.  The negativity $\eta_\theta$ is measured in a single qubit tomography experiment, but $U_\theta$ cannot be directly measured in any such experiment because the composite system contains a qubit defined to be beyond the reach of the experimenter (i.e.\ the bath qubit).  

\subsection{Correlation Alone}
Suppose the composite dynamics are defined by the controlled phase gate, i.e.\ $CZ$, and the sharp operation is defined on the canonical tomography vector $\vec{\tau}$ as 
\begin{equation}
\tau_{i}^{\sharp_\alpha} = \tau_i\otimes\left(U_\alpha \tau_i U_\alpha^\dagger\right)
\end{equation}
with
\begin{equation}
U_\alpha = \alpha\sigma_1 + \sqrt{\left(1-\alpha^2\right)}\sigma_3\;\;,
\end{equation}
where $\sigma_1=\sigma_x$ and $\sigma_3=\sigma_z$ are the standard Pauli operators.  Notice, $U_\alpha$ is unitary if $\alpha\in[0,1]$ with $U_\alpha = H$ if $\alpha = 2^{-1/2}$.  

The Choi representation of the channel would be
\begin{equation*}
\mathbf{C}_\alpha = \begin{pmatrix}
 1 & 0 & 0 & \alpha \sqrt{1-\alpha^2} \\
 0 & 0 & \alpha \sqrt{1-\alpha^2} & 0 \\
 0 & \alpha \sqrt{1-\alpha^2} & 0 & 0 \\
 \alpha \sqrt{1-\alpha^2} & 0 & 0 & 1
\end{pmatrix}
\end{equation*}

The spectrum of $\mathbf{C}_\alpha$ can be written down immediately as
\begin{equation}
\operatorname{spec}(\mathbf{C}_\alpha) = \{1-x_\alpha,-x_\alpha,x_\alpha,1+x_\alpha\}
\end{equation}
where $x_\alpha = \alpha \sqrt{1-\alpha^2}$.  The negativity of this channel $\eta_\alpha$ can be bounded as
\begin{equation}
\alpha\in[0,1]\Rightarrow \eta_\alpha\in\left[0,\frac{1}{6}\right]\;\;,
\end{equation}
with $\eta_\alpha=0$ if $\alpha=0$ or $\alpha=1$ and $\eta_\alpha=1/6$ if $\alpha=2^{-1/2}$.  The negativity $\eta_\alpha$ was already calculated for the case when $U_\alpha=H$ (i.e.\ $\alpha=2^{-1/2}$).

\begin{figure}[th]
\includegraphics[scale=0.52]{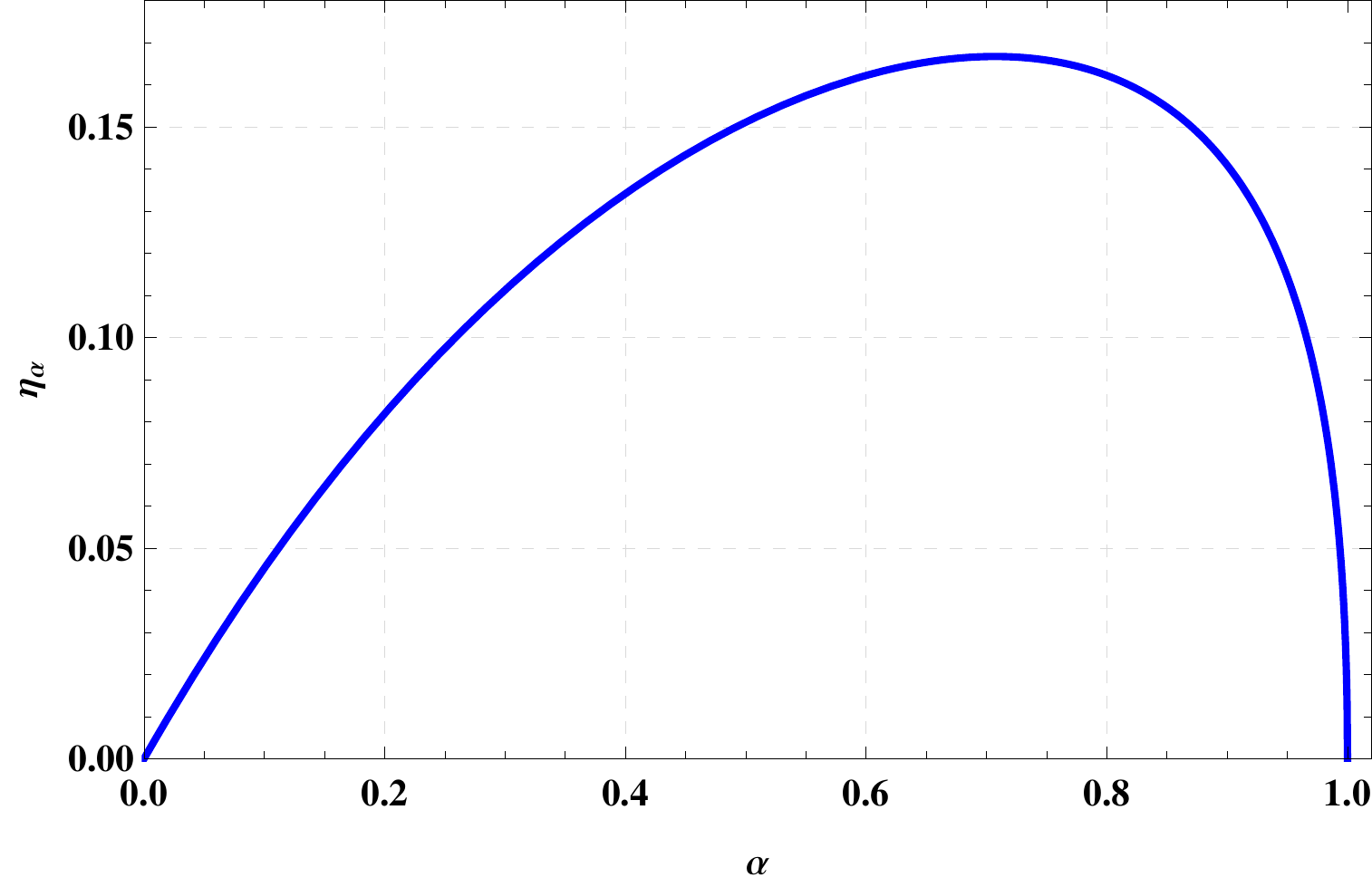}
\caption{The negativity $\eta_\alpha$ can be plotted as a function of $\alpha$ to show the dependency of the negativity on $U_\alpha$.  This example, like the example plotted in Fig.\ \ref{fig:etatheta}, illustrates how the negativity yields information about parameters in the channel definition.  Both $\eta_\alpha$ and $\alpha$ are unitless.}
\label{fig:etaalpha}
\end{figure}

The dependence of $\eta_\alpha$ on $\alpha$ can be plotted (see Fig.\ \ref{fig:etaalpha}). 
Again, the negativity $\eta_\alpha$ can be measured and if the above theoretical definition of the channel is assumed to be true, then $\alpha$ can simply be read off Fig.\ \ref{fig:etaalpha}.  In this example, as in the previous one, measurement of the negativity in a tomography experiment grants the experimenter knowledge about channel parameters that can not be measured directly.

\subsection{Coupling and Correlation Together}

These examples highlight a possible usefulness of the negativity:  measurement of the channel negativity yields information about the bath because the bath influences the channel through the composite dynamics and sharp operation.  Unfortunately, the above examples are artificial in the sense of comparing the experimentally measured negativity to some known analytical definition of the channel (i.e.\ $\mathbf{C}_\theta$ and $\mathbf{C}_\alpha$).  Typically, the experimenter will not have very detailed expectations about the form of the channel in the tomography experiment.  Some assumptions might be made, but it is rare to have a model of the channel complete enough (or in which the experimenter has enough confidence) to do the type of direct comparison between theory and experiment described in the examples.  Typically, the experimenter would be doing tomography experiments precisely to figure out which assumptions about the sharp operation and composite dynamics are reasonable.  Even without precise, confidence-worthy models of the experimental channels, measurement of the negativity will provide information about the composite system.  A good example of this point is the experiment conducted in \cite{Cory2004}, which has already been discussed, where the authors used the negativity (which they called the ``positivity'') to try to determine possible problems with their experimental set-up.  

Consider a slightly more complicated example with the composite dynamics given by $U_\theta$ and the sharp operation from the above example, i.e. consider the channel\
\begin{equation}
\varepsilon(\tau_i) = \trace_B\left(U_\theta\tau_i^{\sharp_\alpha} U_\theta^\dagger\right)\;\;.
\end{equation}
This single qubit channel combines the two above examples and will yield a negativity dependent on both the ``correlation'' (i.e.\ $\alpha$) and the ``coupling'' (i.e.\ $\theta$).  Notice $\theta=0$ and $\theta=2\pi$ yield
\begin{equation}
\mathbf{C}_{0\alpha} = \mathbf{C}_{2\pi\alpha} = \begin{pmatrix}
 1&0&0&1\\
 0&0&0&0\\
 0&0&0&0\\
 1&0&0&1
\end{pmatrix}
\end{equation} 
and $\theta=\pi$ yields
\begin{equation}
\mathbf{C}_{\pi\alpha} = \begin{pmatrix}
 1&0&0&-1\\
 0&0&0&0\\
 0&0&0&0\\
 -1&0&0&1
\end{pmatrix}\;\;.
\end{equation} 
The Choi representations $\mathbf{C}_{0\alpha}$, $\mathbf{C}_{2\pi\alpha}$, and $\mathbf{C}_{\pi\alpha}$ are diagonally dominant and therefore, represent channels with vanishing negativities independent of the value of $\alpha$.

\begin{figure}[tht]
\includegraphics[scale=0.55]{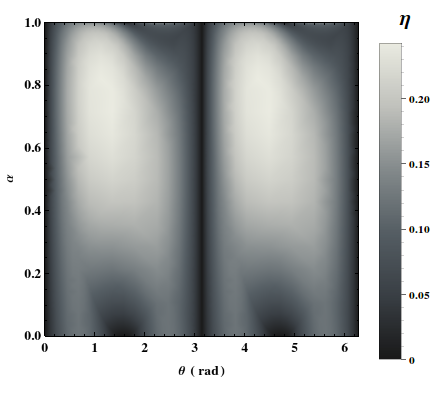}
\caption{The negativity $\eta_{\theta\alpha}$ can be plotted as a function of $\theta$ and $\alpha$ to show the dependency of the negativity on both the correlation and coupling in the channel.  The negativity and $\alpha$ are unitless and $\theta$ has units of radians.}
\label{fig:etathetaalpha}
\end{figure}

The negativity of the channel represented by $\mathbf{C}_{\theta\alpha}$ can be plotted as a function of the full two parameter space (see Fig.\ \ref{fig:etathetaalpha}).  
The maximum negativity $\eta_{\theta\alpha}\approx 0.24$ is achieved around $\alpha=2^{-1/2}$ with either $\theta=\pi/3$ or $\theta=4\pi/3$.  Fig.\ \ref{fig:etathetaalpha} shows the same periodicity about $\theta=\pi$ as Fig.\ \ref{fig:etatheta}, and the maximum of Fig.\ \ref{fig:etathetaalpha} is near the individual maximums found in the single parameter spaces of Fig.\ \ref{fig:etaalpha} and Fig.\ \ref{fig:etatheta}.

The measured value $\eta_{\theta\alpha}$ cannot uniquely identify a location in the two dimensional parameter space plotted in Fig.\ \ref{fig:etathetaalpha}, and this inability is precisely the frustrating limitation of the bath information hidden in the negativity value.    

If the experimentalist were able to measure both the negativity and the initial system-bath correlation, then he would still not be able to draw any conclusions about the causal relationship between the two because the negativity would be confounded by the coupling.  If he were able to measure the negativity and the coupling, then the system-bath correlation would act as the confounder.  In most situations, the experimenter will only be able to measure the negativity and will be unable to understand the causal relationship between those measurements and the preparation procedure (i.e.\ the correlation) or the composite dynamics (i.e.\ the coupling), unless he is able to control for the confounding relationship between them.  For example, in a ``controlled bath'' type experiment, he may be very confident in his understanding of the composite dynamics.  In such a situation, he may indeed be able to understand the causal relationship between the preparation procedure and the negativity.  He could not, however, do so without the extra knowledge/confidence about the composite dynamics.   

The correlation and coupling are confounded in such a way as to limit the experimenter's ability to gain precise information about either one from a measurement of the negativity.  But notice that the measured negativity will limit the possible values of $\theta$ and $\alpha$ to some subset of the total parameter space.  For example, if the experimenter is attempting to use the measured negativity to develop an empirical model of the coupling and correlation, then this smaller parameter space might make the task of comparing numerical simulations to measured data easier.

\subsection{Channel Negativity as a ``Trace Distance''}

Consider a typical quantum information experiment: an experimenter is trying to implement a quantum gate in some new way and would like a measure of the ``correctness'' of the implemented gate.  Suppose the gate to be implemented is the familiar controlled-phase gate $CZ$.  If the experimenter actually implements 
\begin{equation}
CZ^\prime = \begin{pmatrix}
1&0&0&0\\
0&1&0&0\\
0&0&1&0\\
0&0&0&e^{-i \delta}
\end{pmatrix}\;\;,
\end{equation}
where $\delta\in\mathbb{R}\ge 0$, then he would want to know how ``close'' $CZ^\prime$ is to $CZ$.  There are numerous ``gate fidelity'' measures \cite{Wilde2013} \cite{nielsen00}, but a simple choice is the trace distance (derived from the trace norm) given as \cite{Wilde2013}
\begin{equation}
||M-N||_1 = \operatorname{Tr}\left(\sqrt{(M-N)^\dagger(M-N)}\right)
\end{equation}
where $M$ and $N$ are any two operators.  The trace distance is commonly used as a measure of distinguishability between two density matrices \cite{Wilde2013}, but it can be used as a distance measure between $CZ$ and $CZ^\prime$.  The trace distance is not the best choice to determine if the implemented gate behaves as intended (e.g.\ see \cite{nielsen00}), but it has a simple mathematical form and nicely demonstrates a possible use of the negativity of a kind of probe for the system-bath interaction. 

That trace distance is
\begin{equation}
||CZ-CZ^\prime||_1 = 2 \left|\cos\left(\frac{\delta}{2}\right)\right|\;\;.
\end{equation}
This form is very nice, but notice that a measurement of this trace distance would require a two qubit tomography experiment.  The gate actually implemented in the experiment (i.e.\ $CZ^\prime$) would need to be characterized in an experiment which involves a tomography vector of 16 states.

Now consider a negativity measurement on a single qubit channel defined as
\begin{equation}
\varepsilon^\prime(\tau_i) = \trace_B\left(CZ\tau_{i}^\sharp CZ^{\dagger}\right)\;\;,
\end{equation}
where $\tau_{i}^\sharp = \tau_{i}\otimes \left(H\tau_{i} H^\dagger\right)$ is defined on the canonical single qubit tomography vector.  The desired channel will have an expected negativity of $\eta_{CZ} \approx 0.167$ (see Eqn.\ \ref{eqn:Ccz}).

\begin{figure}[th]
\includegraphics[scale=0.55]{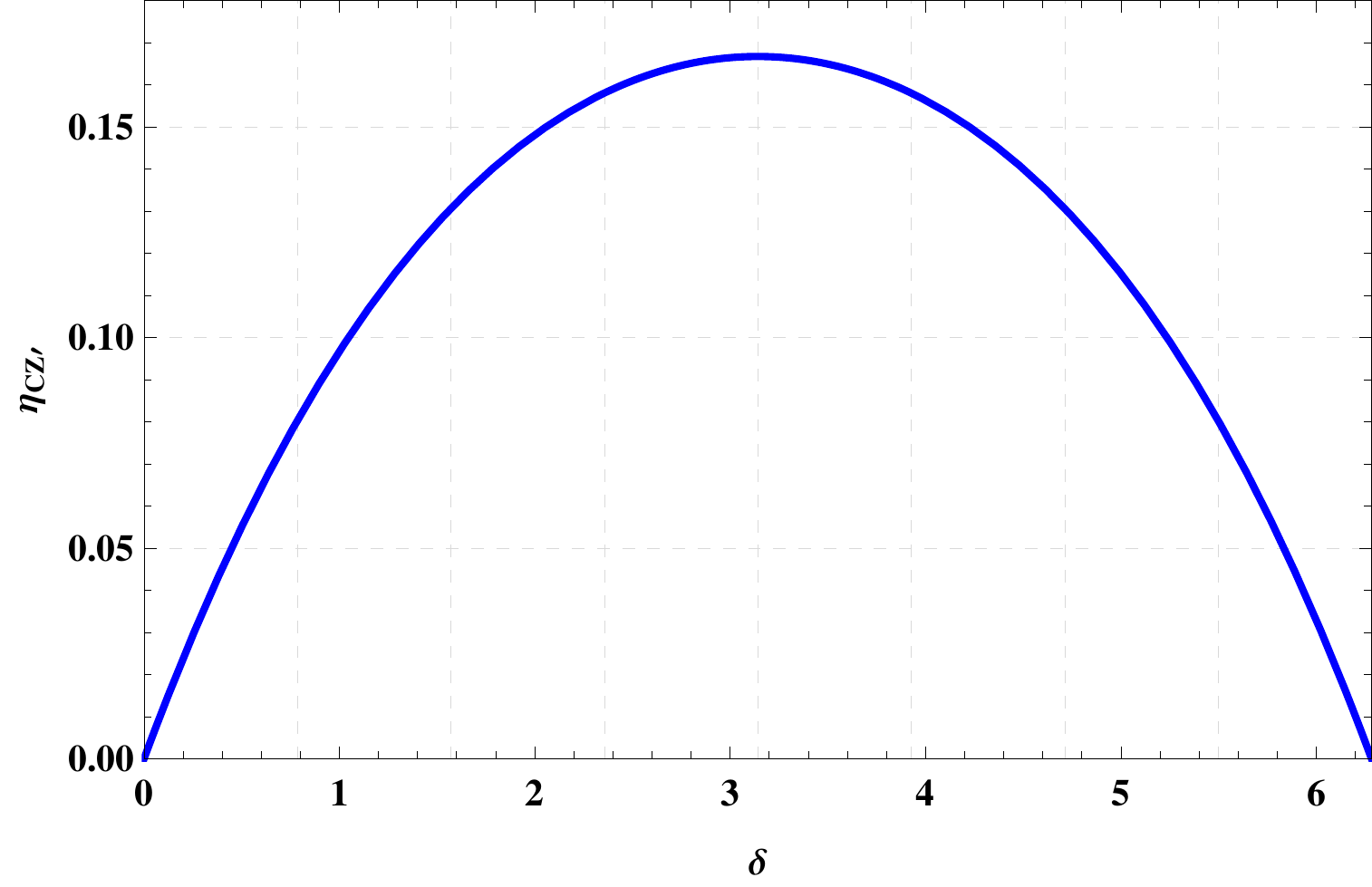}
\caption{The negativity $\eta_{CZ^\prime}$ can be plotted as a function of $\delta$ to show the dependency of the negativity on $CZ^\prime$.  The negativity is maximized at $\delta=\pi$ and has minimums at $\delta=0$ and $\delta=2\pi$.  See the text for discussions of these points and the use of this plot in determining $\Delta$.  Both $\eta_{CZ^\prime}$ and $\delta$ are unitless.}
\label{fig:CZeta}
\end{figure}

The implemented channel will have a Choi representation 
\begin{equation*}
\mathbf{C}_{CZ^\prime} = \begin{pmatrix}
 1 & 0 & 0 & \frac{1}{4} \left(3+e^{i \delta}\right) \\
 0 & 0 & \frac{1}{4}-\frac{e^{-i \delta}}{4} & 0 \\
 0 & \frac{1}{4} \left(1-e^{i \delta}\right) & 0 & 0 \\
 \frac{1}{4} \left(3+e^{-i \delta}\right) & 0 & 0 & 1
\end{pmatrix}.
\end{equation*}
If $\delta = 0$ or $\delta = 2\pi$, then $CZ^\prime = I$ where $I$ is the two qubit identity operator and
\begin{equation}
\mathbf{C}_{CZ^\prime} = \begin{pmatrix}
1&0&0&1\\
0&0&0&0\\
0&0&0&0\\
1&0&0&1
\end{pmatrix}
\end{equation}
which has a vanishing negativity.  If $\delta=\pi$, then $CZ^\prime = CZ$ and the negativity of the implemented channel $\eta_{CZ^\prime}$ will be equal to that of the expected channel, i.e.\ $\eta_{CZ^\prime}=\eta_{CZ}$.  Given $\delta\in[0,2\pi]$, the point $\delta=\pi$ is the only point with this property.  Hence, the quantity 
\begin{equation}
\Delta = |\eta_{CZ}-\eta_{CZ^\prime}|\;\;,
\end{equation}
where $\Delta$ is a ``negativity distance'', can be used to determine how ``far'' the implemented gate $CZ^\prime$ is from the desired gate $CZ$.  Fig.\ \ref{fig:CZeta} shows the negativity of the implemented channel $\eta_{CZ^\prime}$ as a function of the possible values of $\delta$ in $CZ^\prime$ given $\delta\in[0,2\pi]$.
The figure makes it clear that
\begin{equation}
\Delta\in[0,\eta_{CZ}]
\end{equation}
with
\begin{equation}
\Delta=0\Leftrightarrow \eta_{CZ^\prime} = \eta_{CZ}
\end{equation}
and $\Delta = \eta_{CZ}$ if $\delta = 0$ or $2\pi$, but $0< \Delta < \eta_{CZ}$ does not point to a unique $\delta$.  

It should be noted that the trace distance also suffers from the same inability to precisely determine $\delta$.  The point of the trace distance (or the negativity distance) is not to provide a precise characterization of the implemented gate $CZ^\prime$,but rather to provide a sense of distance of the implemented gate from the expected gate.  Either distance measure could be used to determine ``how far'' the implemented gate is from the desired gate, but the experiment to measure $||CZ-CZ^\prime||_1$ differs significantly from the experiment to determine $\Delta$.

\begin{figure}[h!t]
\includegraphics[scale=0.52]{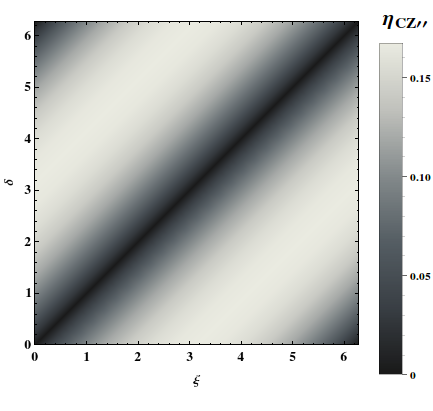}
\caption{The negativity $\eta_{CZ^{\prime\prime}}$ can be plotted as a function of both $\delta$ and $\xi$ to show the dependency of the negativity on $CZ^{\prime\prime}$.  See the text for discussions of how this two parameter space relates to the negativity distance introduced with the single parameter space of $CZ^\prime$ and Fig.\ \ref{fig:CZeta}.  All of the quanities in this plot are unitless.}
\label{fig:CZetaII}
\end{figure}

The negativity distance is still applicable given implemented gates depending on more than one parameter, but the situation becomes expectedly more complicated.  Consider an implemented gate
\begin{equation}
CZ^{\prime\prime} = \begin{pmatrix}
1&0&0&0\\
0&1&0&0\\
0&0&e^{-i \xi}&0\\
0&0&0&e^{-i \delta}
\end{pmatrix}\;\;,
\end{equation}
which depends on two parameters $\{\delta,\xi\}\in\mathbb{R}\ge 0$  and will have some Choi representation $\mathbf{C}_{CZ^{\prime\prime}}$.  Notice, $\delta=\xi$ yields
\begin{equation}
\mathbf{C}_{CZ^{\prime\prime}} = \begin{pmatrix}
 1 & 0 & 0 & e^{i \delta} \\
 0 & 0 & 0 & 0 \\
 0 & 0 & 0 & 0 \\
 e^{-i \delta} & 0 & 0 & 1
\end{pmatrix}
\end{equation}
which has a vanishing negativity.  The negativity of this implemented channel, $\eta_{CZ^{\prime\prime}}$, is plotted as a function of the two parameter space in Fig.\ \ref{fig:CZetaII} given $\{\delta,\xi\}\in[0,2\pi]$.

This plot reveals the difficulty in using a negativity distance, e.g.\
\begin{equation}
\Delta^\prime = |\eta_{CZ^{\prime\prime}} -\eta_{CZ}|
\end{equation}
to characterize the implemented gate $CZ^{\prime\prime}$ in the two parameter space of $\delta$ and $\xi$.  Even a measurement of $\eta_{CZ^{\prime\prime}}=0$ is not unambiguous, because $\delta=\xi\Rightarrow \eta_{CZ^{\prime\prime}}=0$ but $\eta_{CZ^{\prime\prime}}=0$ can mean $\delta=\xi$, $\delta=0$ and $\xi=2\pi$, or $\delta=2\pi$ and $\xi=0$.  This problem will become more pronounced as the parameter space increases in dimension.

The benefit of the negativity distance comes from the manner in which it would be measured.  Complete characterization of the implemented process $CZ^\prime$ (or $CZ^{\prime\prime}$) requires a two qubit process tomography experiment.  Such a complete characterization is required for the use of a distance measure akin to the trace distance.  In contrast, the negativity distance between the implemented and expected channels is calculated using the measured negativity of a single qubit channel; i.e.\ only a single qubit process tomography experiment is needed.  A tomography experiment requires a tomography vector which will have a length of $N^2$ where the number of states in the system is $N=2^n$ given $n$ qubits.  Hence, determining the negativity distance $\Delta$ requires a tomography vector with $2^{2\cdot 2}-2^2=12$ fewer states than the trace distance $||CZ-CZ^\prime||_1$.

The preparation and measurement of 4 states is a significantly simpler experiment then the preparation and measurement of 16 states.  It is in this sense that the negativity distance is ``less complex'' than the trace distance, but notice that the negativity distance and trace distance experiments are not equal in all other aspects.  The negativity measurement of the single qubit channel described above requires the implementation of a specific sharp operation.  The two qubits input into $CZ^\prime$ need to be correlated in a very specific way which might be considered sufficiently complex enough to cause an experimenter to consider the full two qubit process tomography experiment ``easier'' to perform.  The proposed sharp operation has already been given a proposed implementation procedure of projective measurements on a specifically entangled initial composite state \cite{Usha11,Usha11a}, and creating this entangled initial composite state might be very difficult in some experimental set-ups.  This point is a fair criticism to the ``simplicity'' of the negativity distance measurements, but it is not always applicable.  For example, it might be possible that the sharp operation could be implemented without significantly changing the experimental set-up used for the full two qubit process tomography experiment (see \cite{thesis} for examples).

The negativity distance might lead to significantly simpler experiments in the characterization of gates of more than two qubits.  Process tomography becomes much more difficult as the number of qubits increases due to the increased length of the tomography vector.  Negative channels can arise in some situations where the reduced system is correlated to only one of the bath qubits.  Hence, a gate on more than two qubits might be characterized using a negativity distance measurement with a sharp operation between only two of the qubits.  The sharp operation for such an experiment would be no more complex than it would be in an experiment to measure the $\Delta$ defined above, but the single qubit process tomography would be significantly simpler than the ``greater than two qubit'' process tomography that would otherwise be used to characterize the implemented gate.  

\subsection{Determining the Completely Positive Parameter Space}

Consider a ``controlled bath'' type experiment implemented in the lab using the polarization of two maximally entangled photons as qubits.  One photon would act as the reduced system and the other as the bath.  It would be possible to implement a sharp operation very similar to the one presented in Section \ref{sec:Rabi} as a projective measurement on the reduced system photon after applying a rotation to the bath qubit photon \cite{Ralph2010}.  The negativity could then be plotted as a function of time.  Repeating this experiment multiple times with different initial rotation angles for the bath qubit would indicate to the experimenter when the composite dynamics are described by local unitaries (i.e.\ composite unitary evolution that can be written as a tensor product of the reduced system and bath evolutions).  Composite dynamics in local unitary form are the only composite dynamics that are complete positivity for any initial system-bath correlation \cite{Hayashi2003}.  As such, empirically determining when the negativity is zero for a large number of different initial system-bath correlations will allow the experimenter to be reasonably confident of when the composite dynamics are described by local unitaries.  It follows that an experiment which measures the changes in negativity over time for several different initial rotation angles of the bath (i.e.\ $\phi$) can be used to determine when the composite dynamics are in local unitary form without ever knowing anything about the bath dynamics directly.  Such ``controlled bath'' experiments could also be used to understand preparation procedures.

These are specific examples of the most straightforward usefulness of the negativity: mapping the completely positive parameter space.  Complete positivity is, as was discussed in the introduction, a ubiquitous assumption in quantum information theory.  As such, many results rely on the assumption (e.g.\ many quantum error correction techniques, state estimation methods, and definitions of things like ``channel entropy'' rely on the complete positive assumption).  From the definition of the negativity, it can be seen that the negativity is zero if and only if the channel is completely positive.  Thus, determining which experimental parameters yield a vanishing negativity will let the experimenter know where (in parameter space) those completely positive assumptions can be justified.

\begin{figure}[th]
\centering
\includegraphics[scale=0.55]{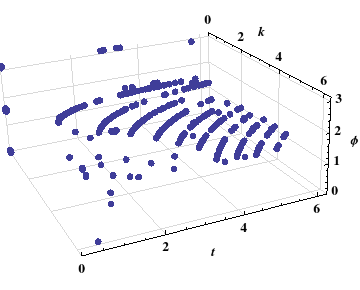}
\caption{The points where the negativity of the channel described in Section \ref{sec:Rabi} are zero in the parameter space of time $t$, coupling constant $k_z$, and the initial rotation angle of the bath qubit $\phi$.  This plot is meant to illustrate the idea of mapping out a completely positive parameter space.  See Section \ref{sec:Rabi} for a discussion of the units of the plotted parameters.}
\label{fig:CPspace}
\end{figure}

For example, suppose some channel has a Choi representation that takes the form
$$
\mathbf{C} = \begin{pmatrix}
1&0&0&x\\
0&0&y&0\\
0&y^*&0&0\\
x^*&0&0&1
\end{pmatrix}\;\;.
$$
(See \cite{thesis} for many examples of such channels.)  The spectrum of $\mathbf{C}$ can be written down as 
$$
\operatorname{spec}\left(\mathbf{C}\right) = \left( 1-\sqrt{xx^*},1+\sqrt{xx^*},-\sqrt{yy^*},\sqrt{yy^*}\right)\;\;.
$$
Notice that $yy^*=0$ and $xx^*\le 1$ are sufficient conditions for this channel to have a vanishing negativity.  These conditions will only be met at specific points in the parameter space of the experiment.  This idea can be illustrated by plotting a few points of vanishing negativity in the 3-dimensional parameter space of $(k_z,t,\phi)$ for the Rabi channel of Section \ref{sec:Rabi} (see Fig.\ \ref{fig:CPspace}).  The assumption of a fixed (Hadamard) rotation in the sharp operation of Section \ref{sec:Rabi} has been dropped to produce this plot.  Also notice that the planes $t=0$ and $k_z=0$ are not plotted because $t=0$ leads to trivial composite dynamics and $k_z=0$ leads to local unitary composite dynamics.  Both situations imply complete positivity and would clutter the plot unnecessarily.  As such, the plot is over the range $\{k_z,t,\phi\}\in(0,2\pi]$.

\section{Conclusions}

The negativity is an extension of the idea of positivity presented in \cite{Cory2004}.  The negativity is real, bounded, and exists for any channel with a Choi representation.  In \cite{Cory2004}, it was used as a way to understand experimental errors in a specific experiment.  The idea was pursued even further in \cite{Wood2009}, when the origin of the negativity in \cite{Cory2004} was studied as a possible by-product of statistical errors in the tomography experiment.  Notice, however, that the negativity can theoretical be non-zero; i.e.\ a vanishing negativity is not always due to error in the implementation of the tomography experiment.  The negativity is non-zero when the composite system exhibits both coupling and correlation between the reduced system and the bath.  As such, these non-zero negativities contain information about the composite dynamics and$/$or the preparation procedure that might be useful to the experimenter.  It can be difficult to determine precisely what of kind information is represented by the negativity measurements in general, but as shown above, the negativity can have a clear connection to the coupling (or correlation) in ``controlled bath'' type experiments.  

The negativity provides information about the bath. Such information is academically interesting, but might have practical use in engineering quantum technologies and understanding the limitations of those technologies.  The negativity can be used to understand when a channel is completely positive.  If complete positivity is assumed to always hold for reduced quantum dynamics, then it is an interesting question to determine what part of the tomography experiments described above are not physical.  If complete positivity is not assumed to always hold, but is desired, then another interesting question would be to determine if useful quantum channels can be restricted to parameter spaces with vanishing negativities even in the presence of imperfect control (i.e.\ imperfect preparations, measurements, etc).

\section{Acknowledgements}

I'd like to thank Keye Martin, Marco Lanzagorta, Johnny Feng, and Tanner Crowder for their help with this idea. 
\bibliographystyle{apsrev4-1}
\bibliography{negativity}
 
\end{document}